\documentclass{svjour3}                     
\smartqed  
\usepackage{graphicx}
\usepackage{mathptmx}      
\usepackage{natbib}
\newcommand{\hzeroplus}{H$^{0}$}		
\newcommand{\honeplus}{H$^{+1}$}		
\newcommand{\carthree}{C~{\footnotesize{III}}}  
\newcommand{\carsix}{C~{\footnotesize{VI}}}  	
\newcommand{\naone}{Na~{\footnotesize{I}}}	
\newcommand{\oxysix}{O~{\footnotesize{VI}}}  	
\newcommand{\oxyseven}{O~{\footnotesize{VII}}}  
\newcommand{\oxyeight}{O~{\footnotesize{VIII}}} 
\newcommand{\oxyfiveplus}{O$^{+5}$}             
\newcommand{\oxysixplus}{O$^{+6}$}             
\newcommand{\oxysevenplus}{O$^{+7}$}             
\newcommand{\ironnine}{Fe~{\footnotesize{IX}}}	
\newcommand{\ironten}{Fe~{\footnotesize{X}}}	
\newcommand{\ironeleven}{Fe~{\footnotesize{XI}}}      
\newcommand{\chips}{{\it{CHIPS}}}		
\newcommand{\copernicus}{{\it{Copernicus}}}	
\newcommand{\chandra}{{\it{Chandra}}}		
\newcommand{\fuse}{{\it{FUSE}}}			
\newcommand{\rosat}{{\it{ROSAT}}}		
\newcommand{\suzaku}{{\it{Suzaku}}}		
\newcommand{\xmm}{{\it{XMM}}}                   
\newcommand{\degree}{$^{\rm{o}}$}		

\journalname{ }

\begin{document}

\title{Revising the Local Bubble Model due to Solar Wind Charge Exchange 
X-ray Emission}

\titlerunning{The Local Bubble vs Solar Wind Charge Exchange}        

\author{Robin L. Shelton
}


\institute{Department of Physics and Astronomy, the University of Georgia,
        Athens, GA 30602}


\date{Received: date / Accepted: date}

\maketitle

\begin{abstract}

The hot Local Bubble surrounding the solar neighborhood has been
primarily studied through observations of its soft X-ray emission.
The measurements were obtained by 
attributing all of the observed local soft X-rays to the bubble.
However, mounting evidence
shows that the heliosphere also produces diffuse X-rays.   
The source is solar wind ions 
that have received an electron from another atom.
The presence of this alternate explanation for locally
produced diffuse X-rays
calls into question
the existence and character of the Local Bubble.
This article addresses these questions.
It reviews the literature on solar wind charge exchange
(SWCX) X-ray production, finding that
SWCX accounts for roughly half of
the observed local 1/4 keV X-rays found at low latitudes.  
This article also makes predictions for the heliospheric \oxysix\ 
column density and intensity, finding them to be smaller than
the observational error bars.   Evidence for the continued
belief that the Local Bubble contains hot gas
includes the remaining local 1/4~keV
intensity, the observed local \oxysix\ column density, and the
need to fill the local region with some sort of plasma.
If the true Local Bubble is half as bright as
previously thought, then its electron density and thermal pressure
are $1/\sqrt2$ as great as previously thought, and its
energy requirements and emission measure are 1/2 as great as previously
thought.    These adjustments can be accommodated easily, and, in
fact, bring the Local Bubble's pressure more in line with
that of the adjacent material.   Suggestions for future work are made.

\end{abstract}



\section{Introduction}
\label{intro}

The Local Bubble (LB) is traditionally thought of as a large ($\sim60$~pc
in radius) bubble filled with hot ($\sim10^6$~K) plasma surrounding
the Solar neighborhood.
Several clouds of warm ($\sim10^4$~K) gas reside within the
Local Bubble.   One of these clouds, the Local Cloud, envelops the
Solar System.

The Local Bubble was discovered in the 1970's 
through 1/4~keV X-ray observations.
Early X-ray instruments detected soft X-rays
from every direction and with B and C band (bandpasses: $\sim70-188$ eV
and $\sim160-284$~eV,
respectively) 
intensities that anti-correlate
with absorbing column density but correlate well with each other.   
These characteristics
suggested that the Earth resides within an X-ray emissive 
bubble which has displaced neutral material in the disk
\citep{mccammon_sanders}.   
That some of the observed X-rays were produced locally and 
not, for example, in the Galactic halo or bulge, has been confirmed by
shadowing studies in which absorbing interstellar clouds are used 
to block X-rays from more distant sources
\citep{burrows_mendenhall,snowden_etal}.
The X-ray intensity observed in the direction of the cloud, minus
the intensity of distant photons that have leaked through the cloud,
indicates the brightness of the local region.
Technically, these studies indicate only that X-rays were produced
``locally'', i.e. somewhere between the satellite and the absorbing 
cloud;  these studies do not
reveal the precise location of the X-ray emitting gas.
Although 
astronomers have long recognized 
the possibility that the Solar System contributed to the observed 
local X-ray intensity, the agreement between 1/4~keV observations taken
more than a year apart \citep{bunner_etal} required a constant
emission source and so favored the interstellar interpretation.
This changed in 1996 when
\rosat\ observed X-rays from the coma of Comet Hyakutake
\citep{lisse_etal}.  These observations unequivocally demonstrated 
that the heliosphere 
contains X-ray sources aside from the Sun and therefore called into 
question the assumption that 
all of the locally produced diffuse X-ray flux
was created by 
the Local Bubble.
The existence and nature of the Local Bubble are now being reevaluated.
This article addresses these issues.
After further describing the historical Local Bubble model and
diffuse Solar System 
X-ray emission, the paper proceeds to a discussion
of whether or not the Solar System can 
entirely replace the
Local Bubble as the explanation for the ``local'' emission
(it cannot) and to a new picture of the Local Bubble which
takes into account the presence of Solar System X-rays. \\

\section{ The Historical Local Bubble Model } 

As mentioned in the introduction, the historical Local Bubble's temperature
is about a million degrees Kelvin.
This temperature was found by comparing observed X-ray band ratios with
models for optically thin hot gas in 
collisional ionizational equilibrium.   
Even the data from the earliest 1/4~keV X-ray telescopes, in which the
Local Bubble emission was not distinguished from 
that of the Galactic halo, matched that of
a $0.9$ to $1.2 \times 10^{6}$~K plasma \cite{mccammon_sanders}.
Later, \rosat\ All Sky Survey data, for which it
was possible to separate the Local Bubble and Galactic
halo contributions, yielded a similar Local Bubble temperature
($T_{LB} = 1.3 \times 10^6$~K, with variation from
$T_{LB} = 1.1$ to $1.9 \times 10^6$~K depending upon direction,
\cite{kuntz_snowden}).   

Warm clouds are buried deep within the bubble.
These have been observed via 
their absorption of light from nearby stars and by analyses of
the material that abuts the heliosphere (see other articles in
this volume for further details).
In addition to these temperature extremes, we expect to find 
intermediate temperature gas in the
transition zones between the clouds and the hot bubble plasma.
Thus, \oxysix, which traces $3 \times 10^5$~K gas was sought.
It was found, both in a statistical analysis of dozens of
\copernicus\ column density measurements \cite{shelton_cox}
and, also in analyses of 
\fuse\ data for stars within about a hundred parsecs of the Sun
\cite{oegerle_etal,savage_lehner}.
This distance range is thought to be similar to the Local Bubble's radius,
which was determined
from data on a molecular cloud residing just inside the bubble boundary
\cite{snowden_mccammon_verter}.
Once the radius was known for one direction, the radii
for other directions were estimated from the observed intensities
in those directions and
the assumption that the temperature and density are the same
in all directions.
By this method, the distance from the Sun to the Local Bubble's
periphery was found to vary from about 40~pc
to slightly more than 100~pc \cite{snowden_etal_98}.
Measurements of \naone\ column densities toward nearby stars indicate that
a larger cavity, called the Local Cavity, also surrounds the
solar neighborhood.   This region of exceptionally low
volume density varies in radius from $\sim40$ to $\sim 200$~pc
\cite{lallement_etal}.   
The Local Bubble fits easily within the Local Cavity.

Several models have been proposed to explain the Local Bubble,
including a single supernova explosion \cite{cox_anderson,edgar},
multiple supernova explosions \cite{avillez_breitschwerdt},
and a bubble that has broken out of its nascent molecular cloud
\cite{breitschwerdt_schmutzler}.   Multiple supernova models
are currently favored because of the large input energy
requirement to blow such a large and energetic bubble and
because the observed 1/4 keV X-ray to \oxysix\ and \carthree\
ratios disallow the blow-out model \cite{welsh_etal,shelton,oegerle_etal}.

It is noteworthy that time variability was seen in the \rosat\ data.
The 
\rosat\ All Sky Survey scans overlapped each other.   By comparing
data taken during one orbit with that taken during the next,
it was possible to see that the flux level varied with time
in addition to position on the sky.   The excess fluxes, called 
``long term enhancements'' were nominally removed from the
\rosat\ All Sky Survey data before the data were used to 
measure the Local Bubble's temperature and brightness.
The long term enhancements were unexplained and troubling.
But, after X-rays were seen from 
Comet Hyakutake, \cite{cravens} suggested that solar wind X-rays 
could have caused the long term enhancements in the \rosat\ survey.
A more troubling notion is the possibility that
solar wind X-rays may cause additional contamination.

\section{Charge Exchange in the Solar System }

\subsection{Solar Wind Charge Exchange X-rays}
\label{solarsystemxrays}

The physical mechanism by which Comet Hyakutake emitted X-rays also acts
throughout the Solar System.
In this mechanism, highly charged ions in the solar wind
collide with and receive electrons from other atoms.
The electrons transfer into high $n$ 
levels in the solar wind ions, then they radiatively decay 
by emitting one or more photons. 
For example,
\oxysevenplus\ + \hzeroplus\ $\rightarrow$ \oxysixplus$^*$ + \honeplus\ 
$\rightarrow$ \oxysixplus\ + \honeplus + {\rm{photons}}.   
During the radiative decay, the electron may pass through several excitation
levels.  The last transition, the $n \geq 2$ to $n = 1$ transition, 
will produce an X-ray photon. 
For example, in \oxysixplus, each $n = 2$ to $n = 1$ transition produces 
a $\sim570$ eV X-ray photon.

The charge transfer interactions, called Solar Wind Charge Exchange (SWCX), 
are expected to occur between solar wind
ions and neutral interstellar atoms that have drifted into the heliosphere
(called heliospheric SWCX),
and between solar wind ions and
material in the Earth's upper atmosphere (called geocoronal SWCX).
SWCX interactions with material flowing into the heliosphere 
were found to outnumber SWCX interactions with material in the
Earth's atmosphere \cite{robertson_cravens}.
The solar wind is non-isotropic and time varying on long 
and short 
time scales.    
The longest timescale is $\sim11$~years and is associated with
the Solar activity cycle.   The shortest timescale is 
on the order of an hour and is caused by Coronal Mass Ejections.

Multiple research groups have modeled the SWCX emission of soft
X-ray photons.
Concentrating on the 1/4~keV X-ray band, and accounting for the
state of the Solar Cycle and the sight line geometries during the
\rosat\ All Sky Survey, 
\cite{robertson_cravens}
calculated the heliospheric and geocoronal SWCX contributions to
the 1/4~keV flux observed by \rosat.    
Their SWCX emission map (Figures 9 and 10 in their paper)
is markedly non-isotropic,
with two dim regions centered on 
$\ell \sim 220$\degree, $b \sim -45$\degree\ and
$\ell \sim 130$\degree, $b \sim 0$\degree\ and with an extended bright
region whose intensity peaks near $\ell = 240$\degree, $b = 45$\degree.
Their conservatively estimated intensity of 1/4~keV emission accounts for 
about half of
the diffuse 1/4~keV X-ray intensity seen in the Galactic plane and
thus accounts for about half of the intensity previously attributed
to the Local Bubble in the Galactic plane.
The comparison at higher latitudes is more complicated due to the
Galactic halo contribution.
Subsequently, 
\cite{koutroumpa_etal_06} modeled the heliospheric
\oxyseven\ 
SWCX emission 
at $\sim570$ eV and
estimated that 
as many as 0.8 to 3.4 photons cm$^{-2}$ s$^{-1}$ sr$^{-1}$ could  
result from SWCX in the slowly varying solar wind,
depending on the
stage of the Solar cycle and the viewing angle.   
This range is of the order of what is expected from the Local 
Bubble. 
Note that the \oxyseven\ triplet at $\sim$570~eV 
lies outside \rosat's 1/4 keV band, but within \rosat's 3/4 keV
band, as well as \xmm, \chandra, and \suzaku's bandpasses.
The Local Bubble is much dimmer in the 3/4 keV
band than the 1/4 keV band. Therefore the LB's \oxyseven\ spectral features 
are not expected to be bright and the SWCX photons may account for
a relatively larger fraction of the observed intensity.

In addition to the heliospheric and geocoronal SWCX components, 
there seems to be
an even brighter component which may be associated with Coronal Mass Ejections
(CMEs).   Large CMEs discharge $\sim 10^{11}$ to $\sim10^{13}$~kg
of ionized material non-isotropically into the heliosphere over 
periods of hours.  As this material flows outwards through the
heliosphere, it is subject to charge exchange that leads to X-ray emission.
This is the conclusion drawn by \cite{koutroumpa_etal_07} and used to 
explain the relatively high \oxyseven\ intensities seen in
\chandra\ and \xmm\ shadowing observations.   
In one
example, \xmm\ observed a pair of shadowing sight lines in
2002, during the solar maximum when CMEs are most common.
\suzaku\ observed the sightlines in 2006, during Solar minimum
when CMEs are least common.   Between the 2002 and 2006 observations,
the ``locally produced'' \oxyseven\ intensity dropped from
$6.1^{+2.8}_{-3.0}$ to $0.4 \pm 1.1$ photons cm$^{-2}$ s$^{-1}$ sr$^{-1}$
\citep{henley_shelton}.
Not only does a CME provide enough material to explain the
large intensity difference, but, as
\cite{henley_shelton} point out, it also provides an
explanation for why there was no anomaly in the solar wind
proton data recorded by {\it{ACE}} SWEPAM near the time of the
\xmm\ observation.   {\it{ACE}} monitors the solar wind $\sim0.01$~AU
upstream from the Earth,
and it would not have noticed a CME that
did not pass through the satellite.

\subsection{Solar Wind Charge Exchange \oxysix}
\label{solarsystemovi}

Although SWCX is an important contaminant to the 
observations of diffuse X-rays, it is not a significant
contaminant to the \oxysix\ observations. 
This can be verified by the following estimates.
In the first estimate, 
the \oxysix\ column density, $N_{OVI}$, is found from 
$\int n_{OVI} dl$, where
the volume density of \oxysix\ ions, $n_{OVI}$, can
be estimated from the volume density of solar wind protons, $n_p$,
the oxygen-to-hydrogen ratio, $[O/H]$, and the
fraction of solar wind oxygen atoms in the \oxyfiveplus\ ionization state.
The solar wind proton density decreases with distance from
the Sun, $r$, as $n_p = n_{p1} (r_1/r)^2$, where $r_1$ is
1 AU and $n_{p1}$ is the proton density at $r_1$.
The proton density at $r_1$ can be found from
the relationship between the proton density, proton flux, and
solar wind velocity 
(i.e. $n = F/v$), using
\cite{schwadron_cravens}'s estimates for the
solar wind proton flux at $r_1$,
($F_1 = 2 \times 10^8$~cm$^{-2}$ s$^{-1}$), 
and solar wind velocity (fast wind: 810~km s$^{-1}$,
slow wind: 442~km s$^{-1}$, simple average: 626~km s$^{-1}$).
Thus $n_{p1} = 3.2$~cm$^{-3}$.  
\cite{schwadron_cravens}'s estimate for the
oxygen-to-hydrogen ratio in the solar wind is also used.   The ratio
varies from 1/1780 for the slow wind to 1/1550 for the fast wind.
The simple average of these numbers yields 
$[O/H] \sim 6.0 \times 10^{-4}$ oxygens per hydrogen.
All of the solar wind oxygen atoms are ionized at the \oxysixplus\
level or above when they leave the Sun
\citep{schwadron_cravens}.   Some of these ions will charge
exchange to the \oxyfiveplus\ level while in the Solar System,
so the upper limit on the \oxyfiveplus\ to oxygen ratio along
a given sightline is 1.0.
As a result, the upper limit on $N_{OVI}$ is
$1.9 \times 10^{-3}$~cm$^{-3}$ $\int (r_1/r)^2 \ dl$.
The integral is equal to 0.99~AU, for the 
simplest path, which begins at the Earth and proceeds
directly away from the Sun to the heliopause located about 100~AU downstream.
Thus, the estimated $N_{OVI} \stackrel{<}{\sim} 
2.9 \times 10^{10}$~cm$^{_2}$.   
This value is much smaller than
the error bars on some of the smallest 
column densities observed by \fuse, such as
$N_{OVI}=2.4\pm{1.4} \times 10^{12}$~cm$^{-2}$ \citep{savage_lehner}, and
$N_{OVI}=0.3\pm{2.3} \times 10^{12}$~cm$^{-2}$ \citep{oegerle_etal}.

The {\underline{intensity}} of SWCX-induced \oxysix\
resonance line photons (1032, 1038~\AA) has not been estimated
by \cite{robertson_cravens,koutroumpa_etal_06,koutroumpa_etal_07}.
However, an estimate can be made from the information in 
\cite{koutroumpa_etal_06}.   They predicted the intensity
of heliospheric SWCX 72 and 82 eV photons, which are
produced by \oxyfiveplus\ ions undergoing
transitions from the $n=3$ shell to the $n=2$ shell.   
Note that some of these transitions will place the 
electron into the {\it{p}} subshell, while others will place it
into the {\it{s}} subshell.   Those left in the {\it{p}} subshell will then
undergo the $2p\ ^2P_{3/2}$ to $2s\ ^2S_{1/2}$ or the
$2p\ ^2P_{1/2}$ to $2s\ ^2S_{1/2}$ transitions that
yield 1032 and 1038 \AA\ \oxysix\ resonance line photons.
Thus, the upper limit on the
\oxysix\ resonance line intensity due to heliospheric SWCX is set by
\cite{koutroumpa_etal_06}'s predictions for the 72 and 82 eV photons.
The greatest intensity of these
photons shown in their Figure 1 maps is 
$3.3 \times 10^{-9}$ ergs s$^{-1}$ cm$^{-2}$ sr$^{-1}$.
Thus, the \oxysix\ resonance line intensity can be
estimated as $<$ 26 photons s$^{-1}$ cm$^{-2}$ sr$^{-1}$.
Even if this upper limit were to be elevated by a factor of
several due to a CME along the line of sight, it would be
much smaller than the $1\sigma$ error bars on the
\fuse\ observations of the local \oxysix\ resonance line
intensity ($\sim200$ photons s$^{-1}$ cm$^{-2}$ sr$^{-1}$
for each of the resonance lines, \cite{shelton}), 
showing that that SWCX is not an important factor when
evaluating \oxysix\ emission observations.

\section{ Is There a Hot Local Bubble After All ? }

There is a hot Local Bubble.
\cite{robertson_cravens_snowden} found that maps of the 1/4~keV X-ray
sky are still bright,
even after the SWCX intensity is subtracted.  
Especially noteworthy is that there are regions in the Galactic
plane with non-zero net intensity
(countrate $\sim200\times 10^{6}$~counts s$^{-1}$ arcmin$^{-2}$).
Due to the high opacity of the Galactic disk, only soft X-rays
that are made locally and a small fraction of the X-rays
that come from very bright non-local sources will be observed
at Earth.   Thus, low latitudes X-rays are almost
entirely attributable to the local region and the mapped intensity
of low latitude X-rays remaining after SWCX X-rays have been subtracted 
is almost entirely attributable to the Local Bubble.
The story becomes more complicated when one considers the
3/4~keV band, or one of its most prominent features, the \oxyseven\
triplet.  According to \cite{koutroumpa_etal_07}, it is possible that
SWCX might account for as much as $100\%$ of the measured local \oxyseven\ 
intensity. However, given the inherent uncertainty
in the SWCX \oxyseven\ estimates, there is
room for both the Local Bubble and Solar Wind Charge Exchange.
Also, even if the majority of local \oxyseven\ photons resulted from
SWCX events, the Local Bubble would not be doomed.  The Local Bubble
is not thought to be especially bright in 3/4~keV X-rays
(the local 3/4 keV countrate 
is only $\sim1/15$ as bright as the local 1/4~keV countrate 
\cite{snowden_mccammon_verter}), which makes it easier for
a contaminant such as SWCX to produce a large fraction of the
observed countrate.   

One of the arguments for the continued belief in the 
Local Bubble is that the \oxysix\ column density 
found between the Earth
and nearby stars \cite{jenkins,oegerle_etal,savage_lehner} 
implies the presence of hot gas and cannot be explained by SWCX.  
The \oxysix\ ion is a tracer of $\sim3 \times 10^5$~K gas.
Gas of this temperature is thermally unstable; 
it does not remain at this temperature for long periods 
unless hotter gas is present to resupply it.  
Therefore, it is 
primarily located in transition zones between hotter and cooler gas.
Its existence within the local region has been found by     
\copernicus, which observed measureable quantities of \oxysix\ ions 
on 2 sight
lines terminating within 90~pc of the Sun and several more sight lines
terminating within 200~pc of the Sun \cite{jenkins}, and later
by \fuse, which observed \oxysix\ ions
and on dozens of sight lines with similar distance ranges 
\cite{oegerle_etal,savage_lehner}.
It should be noted that the original \fuse\ work has been 
questioned by \cite{barstow}, who have
reprocessed the original data using 
a more recent version of the \fuse\ data processing pipeline.
They found that some of the previously observed
\oxysix\ absorption features may be artifacts of the older data 
processing, while other features might be circumstellar features
due to the white dwarf targets.  
Nonetheless, \cite{barstow}  did not contradict all of the
earlier \fuse\ detections.   They
found interstellar \oxysix\
on several close sight lines, including WD0004+330, WD0354-368, 
WD2000-561, and WD2331-475.   Because they found far less \oxysix\
within the LB's radius than had \cite{oegerle_etal,savage_lehner},
they concluded that the interstellar \oxysix\ only resides near
the LB's periphery, and not within the LB as \cite{oegerle_etal}
had concluded.
The \cite{barstow} reanalysis does not
extend to the \copernicus\ data.   None of the \copernicus\
targets were white dwarfs; all of the nearby targets were B stars.
Furthermore, the orginal work by \cite{jenkins,jenkins_b} ruled out 
circumstellar orgins for the \oxysix\ found in the \copernicus\ data set.
As shown in Section~\ref{solarsystemovi},
the observed \oxysix\ ions cannot be attributed
to the solar wind ions or SWCX, 
and so the only choice is that they
are associated with hot interstellar gas in the local region, i.e. 
in the Local Bubble.

Another argument for the Local Bubble's existence is that 
the Local Cavity must be occupied.   
If there is no Local Bubble to balance the thermal and magnetic
pressure of the material outside the Local Cavity, then
the material outside the cavity would
expand inwards at the sound speed.   Similarly, the   
Local Cloud that surrounds the Solar System and 
other clouds 
that are also embedded in the Local Bubble's volume would
also expand at the sound speed.
Such expansion, with predicted velocities of about 15 km s$^{-1}$,
has not been observed \cite{slavin}.   
It might be argued that it is the total pressure (including
magnetic and cosmic ray pressures) rather than
the thermal pressure, alone, which governs the dynamics.
Therefore, it may be possible for the total pressure in the Local Cavity to
balance that in the embedded clouds and the surrounding material
if the non-thermal pressure in the Local Cavity is unusually large.
In response, one could point out that
some researchers find low magnetic field 
strengths ($B = 0.7\ \mu$G, \cite{spangler}) 
although other researchers find high magnetic
field strengths ($B = 8\ \mu$G, \cite{andersson_potter})
on lines of sight terminating within 200~pc of the Sun.   

Also note that during the conference, some participants 
added additional reasons
to this list.  
Steve Snowden pointed out that there is good agreement between
the Wisconsin, HEAO~1, and \rosat\ 1/4 keV band data
\citep{snowden_etal_95}, yet these  
surveys were made at different times and with different angles 
between the target and the Sun.   Such agreement is not possible
if most of the X-rays derive from a strongly time and look-angle dependent
source such as the solar system.   Furthermore,
the anticorrelation between neutral hydrogen column density
and soft X-ray intensity cannot be explained by SWCX.
For a debate of the arguments for and against the existence of the
Local Bubble, the reader is refered to the session
report in this volume \cite{shelton_session}.

\section{ Revising the Local Bubble } 

If, as \cite{robertson_cravens} conservatively estimated, 
half of the local 1/4 keV X-rays 
are due to SWCX and half are due to the
hot Local Bubble, then the Local Bubble must be considerably weaker
than previously thought.   
Its electron density and thermal pressure would be reduced
by a factor of $1/\sqrt{2}$ from their previous values 
to $n_e \sim 0.005$ cm$^{-3}$
and $P_{th}/k \sim 11,000$ K~cm$^{-3}$, respectively, if the
gas temperature and path-length remain unchanged.
These reductions are comforting, because the previous pressure 
estimate ($P_{th}/k \sim 15,000$ K cm$^{-3}$, \cite{snowden_etal_98}) was 
far greater than that of the embedded warm clouds 
($P_{th}/k \sim 2000$ to $3000$~K cm$^{-3}$).

The reductions may also help to explain the dearth of 
\ironnine, \ironten, and \ironeleven $\sim72$~eV photons 
observed by 
the X-ray calorimeter \cite{mccammon_etal}
and 
\chips\ \cite{hurwitz_etal}.  
This is a muddy
issue, because \chips\ observed a significantly lower intensity
than did the calorimeter and because preferential depletion of 
gas phase iron relative to other elements can partly explain
the dearth of photons observed by
\chips.   Nonetheless, if SWCX caused half of the
observed local 1/4~keV emission but did not contribute \ironnine, 
\ironten, or \ironeleven\ photons, then the LB's emission
measure would be half that previously expected 
(thus half of $0.0038$~pc$^{-1}$~cm$^{-6}$ expected 
for the \chips\ observations assuming a temperature of $10^{6.1}$~K, 
\cite{hurwitz_etal})
bringing it well within the $95\%$ upper limit contour found from
their observations ($0.0036$~pc$^{-1}$~cm$^{-6}$).

A weaker Local Bubble requires $1/\sqrt{2}$ less 
initial input energy.   Rather than 
a few supernova explosions, perhaps
only one or two would be sufficient for evacuating and heating the bubble.
Another consequence of considering a significant SWCX contribution 
is the possibility that 
it has 
skewed the LB temperature measurements.
The present observational evidence tells us the
shape of the combined LB$+$SWCX spectrum.   The SWCX spectrum must
be subtracted in order to find the true shape of the LB spectrum.
Therefore, the Local Bubble could be hotter or cooler than previously 
believed.  
There may already be observational hints of this effect.   
For example, \xmm\ and \suzaku\ observed the same cloud shadow
4 years apart.   The \xmm\ data was later found to be contaminated
by a SWCX CME.    Before that was known, the 
\xmm\ data, in combination with \rosat\ $1/4$~keV data, was used to
find the temperature of the local plasma 
(i.e. the temperature of the SWCX and LB spectra, combined; 
$\log(T) = 6.06^{+0.02}_{-0.04}$~K, \cite{henley_shelton_kuntz}).
But, the \suzaku\ data, which is not known to be contaminated by
a CME, in combination with the same \rosat\ data,
yielded $\log(T) = 5.96^{+0.05}_{-0.04}$~K, \cite{henley_shelton}.

Lastly, as a result of SWCX contamination, some LB models that had
previously been found to be in conflict with the observations may need
to be reevaluated.   For example, the break-out model
(in which the Local Bubble gas was heated and then expanded so fast that the
ions adiabatically cooled faster than they recombined, 
\cite{breitschwerdt_schmutzler}) had been
eliminated on the grounds that its ratios of 1/4~keV to \oxysix\
emission, \carthree\ emission, and \oxysix\ absorption column
density were higher than the observations indicated.    
But, if the LB's soft X-ray intensity were to be revised 
downwards significantly, then the break-out model 
would require reconsideration.

\section{How We Can Make Progress} 

Continued work on SWCX estimates and the uncertainties on those
estimates will enable continued progress in determining 
how best to apportion the observed local X-ray intensity between
the SWCX and LB components.
Continued work on estimating the SWCX 1/4 to 3/4 keV spectrum
and its subtraction from the observationally derived local
spectrum will enable revised estimates of the LB's spectrum
and hence the plasma's temperature.   But, such a statement
must be qualified, because
1.) some estimates already exist \cite{cravens} although
the spectral subtraction has yet to be done; and
2.) if the SWXC contribution is large, then the uncertainty
in the LB spectrum will be very large.

Clever strategies are needed in order to set observational
constraints on the SWCX intensity.  Perhaps these strategies
will use the temporal and spatial variations in the model SWCX
emission.   Or perhaps, they will use a shadowing strategy to
observe the SWCX spectrum without ``contamination'' by Galactic 
X-rays.
Observations in the 1/4 keV band would be very useful, although
they may not be attainable with current instruments.
Concentrating on this band is necessary because 
the observed local X-ray spectrum peaks in the 1/4 keV band, and as a result,
observations in higher bands alone are not sufficient for
constraining fits to spectral models.
In the future, very high spectral resolution observations
will be useful.  
Few to several-eV resolution is necessary for 
unambiguous measurements of 
atomic transitions in the cascade from high $n$ levels.
Such measurements will include the determination of the ratio of
intensities in the \oxyseven\ ``triplet''
at 561, 569, and 574~eV and 
measurements of the weak \oxyeight\ Ly$\delta$ line at 836~eV
and the weak \carsix\ Ly$\gamma$ line at 459~eV predicted by 
\cite{wargelin_beiersdorfer_brown}.

\begin{acknowledgements}

I would like to thank all of the members of my splinter
group 
for their lively and informative contributions to the 
discussions of the Local Bubble
and to thank the conference organizers for enabling these exchanges.
I would also like to acknowledge helpful comments on the
manuscript from Jeff Linsky,
informative conversations with
Dave Henley, an excellent explanation of solar wind charge exchange
by Tom Cravens, and funding through the NASA LTSA grant NNG04GD78G.

\end{acknowledgements}


\end{document}